
\def\Z_2{{Z \!\!\! Z}_2}
\def\[{[\![}
\def\]{]\!]}
\def\q{\hat q}

\magnification 1200
\hfill TWI-94-29
\bigskip\bigskip\bigskip\bigskip
\noindent
{\bf QUANTIZATION OF THE LIE ALGEBRA SO(2N+1) AND  OF THE
LIE SUPERALGEBRA OSP(1/2N) WITH PREOSCILLATOR GENERATORS}\footnote*{An
invited talk at the Meeting of the FNRS Contact Group on Mathematical
Physics, Universit\'e Libre de Bruxelles, October 1994}

\vskip 28pt

\centerline {T. D. Palev\footnote{**}{Permanent address: Institute for Nuclear
Research and Nuclear Energy, 1784 Sofia, Bulgaria; E-mail
palev@bgearn.bitnet}}

\noindent
\centerline{{\it Department of Applied Mathematics and Computer Science,}}
\centerline{{\it University of Ghent, Krijgslaan 281-S9, B-9000 Gent, Belgium}}
\centerline{{\it e-mail: Tchavdar.Palev@rug.ac.be}}

\vskip 28pt
\centerline{\bf Abstract}
\bigskip

The Lie algebra $so(2n+1)$ and the Lie superalgebra $osp(1/2n)$
are quantized in terms of $3n$ generators, called preoscillator
generators.  Apart from $n$ "Cartan" elements the preoscillator
generators are deformed para-Fermi operators in the case of
$so(2n+1)$ and deformed para-Bose operators in the case of
$osp(1/2n)$. The corresponding deformed universal enveloping
algebras $U_q[so(2n+1)]$ and $U_q[osp(1/2n)]$ are the same as
those defined in terms of Chevalley operators. The name
"preoscillator" is to indicate that in a certain representation
these operators reduce to the known deformed Fermi and Bose
operators.

\vskip 28pt
\noindent
{\bf 1. Preoscillator realization and oscillator representations of
osp(2n+1/2m) and of some of its subalgbras [1]}

\bigskip
The Lie superalgebra $osp(2n+1/2m)\equiv B(n/m)$ can be defined
as the set of all matrices of the form (T=transposition)[2]:
$$
\pmatrix{ a     & b     & u     & x & x_1 \cr
          c     &-a^T   & v     & y & y_1 \cr
	 -v^T   &-u^T   & 0     & z & z_1 \cr
	  y_1^T & x_1^T & z_1^T & d & e   \cr
	 -y^T   &-x^T   &-z^T   & f &-d^T \cr},\eqno(1)
$$
\smallskip
\noindent
where $a$ is any $(n\times n)$ matrix; $b$ and $c$ are skew symmetric
$(n\times n)$ matrices; $d$ is any $(m\times m)$; $e$ and $f$ are
symmetric $(m\times m)$ matrices; $u$ and $v$ are $(n\times 1)$ columns;
$x, x_1, y, y_1$ are $(n\times m)$ matrices and $z, z_1$ are
$(1\times m)$ rows.

The even subalgebra consists of all matrices with
$$x=x_1=y=y_1=z=z_1=0,  $$

\noindent
namely

$$
\pmatrix{ a     & b     & u   & 0 &  0  \cr
          c     &-a^T   & v   & 0 &  0  \cr
	  -v^t  &-u^T   & 0   & 0 &  0  \cr
	  0     & 0     & 0   & d & e   \cr
	  0     & 0     & 0   & f &-d^T \cr}\eqno(2)
$$

\smallskip
\noindent
and it is isomorphic to the Lie algebra $so(2n+1)\oplus sp(2m)$.
The odd subspace consists of all matrices

$$
\pmatrix{   0   & 0     & 0     & x & x_1 \cr
            0   & 0     & 0     & y & y_1 \cr
	    0   & 0     & 0     & z & z_1 \cr
	  y_1^T & x_1^T & z_1^T & 0 & 0   \cr
	 -y^T   &-x^T   &-z^T   & 0 & 0  \cr}.\eqno(3)
$$

The product (= the supercommutator) is defined on any two
homogeneous elements $a,b$ from $osp(2n+1/2m)$ as

$$\[a,b\]=ab-(-1)^{deg(a)deg(b)}ba.\eqno(4)   $$

An important role in the construction plays the
$2(n+m)-$dimensional ${\bf Z}_2-$graded   subspace
$G(n/m)$ consisting of all matrices

$$
\pmatrix{ 0     & 0     & u     & 0 & 0   \cr
          0     & 0     & v     & 0 & 0   \cr
	 -v^T   &-u^T   & 0     & z & z_1 \cr
	  0     & 0     & z_1^T & 0 & 0   \cr
	  0     & 0     &-z^T   & 0 & 0  \cr}.\eqno(5)
$$

Label the rows and the columns with the indices
$$
A,B=-2n,-2n+1, \ldots,-2,-1,0,1,2,\ldots,2m
$$
and let $e_{A,B}$ be a matrix with 1 at the intersaction of the
$A^{th}-$row and the $B^{th}-$column and zero elsewhere.
Choose as a basis in $G(n/m)$ the following  elements (matrices):

$$\eqalign{
& C_i^-(1)\equiv B_i^- =\sqrt{2} (e_{0,i}-e_{i+m,0}),
\;i=1,\ldots,m, \cr
& C_i^+(1)\equiv B_i^+ =\sqrt{2} (e_{0,i+m}+e_{i,0}), \cr
& C_j^-(0)\equiv F_j^- =\sqrt{2}(e_{-j,0}-e_{0,-j-n})
,\;j=1,\ldots ,n,  \cr
& C_j^+(0)\equiv F_j^+ =\sqrt{2}(e_{0,-j}-e_{-j-n,0}).
}\eqno(6)
$$

\noindent
The elements $C_i^\pm(1)$ are odd and $C_j^\pm(0)$ - even.
We call all these elements creation and annihilation operators (CAOs)
of $osp(2n+1/2m)$.

\bigskip
{\it Proposition 1} [1]: The LS $osp(2n+1/2m)$ is generated from
its creation and annihilation operators.

\smallskip
It turns out that already the supercommutators of all CAOs
give the lacking basis elements. Therefore

$$
lin.env.\{\[C_i^\xi (\alpha),C_j^\eta (\beta)\],\;
C_k^\varepsilon (\gamma)|
 \forall i,j,k; \;
\xi ,\eta ,\varepsilon =\pm ; \; \alpha ,\beta
,\gamma\in \Z_2 \}
= osp(2n+1/2m).\eqno(7)
$$

\noindent
Hence any further supercommutator between

$$\[C_i^\xi (\alpha),C_j^\eta (\beta) \] \quad {\rm and}
\quad C_k^\varepsilon (\gamma)$$

\noindent
is a linear combination of the same
type of elements. The more precise computation gives:

$$
\eqalignno{
& \[\[C_i^\xi (\alpha),C_j^\eta (\beta) \],C_k^\varepsilon (\gamma) \]=
2\varepsilon^{\gamma} \delta_{\beta \gamma}\delta_{jk}
\delta_{\varepsilon,-\eta}C_i^\xi (\alpha)
-2\varepsilon^\gamma (-1)^{\beta \gamma}\delta_{\alpha \gamma}
\delta_{ik}\delta_{\varepsilon,-\xi}
C_j^\eta (\beta ), & (8) \cr
} $$
\smallskip
\noindent
where $\; \xi ,\eta ,\varepsilon =\pm ,\; \alpha ,\beta
,\gamma\in \Z_2$ and $i,j,k $ take all possible values.

\bigskip
In the case $\alpha =\beta =\gamma =0$ (8) reduces to

$$[[F_i^\xi ,F_j^\eta ],F_k^\varepsilon ]=
2\delta_{jk}
\delta_{\varepsilon,-\eta}F_i^\xi -
2\delta_{ik}\delta_{\varepsilon,-\xi}
F_j^\eta ,  \eqno(9)  $$

\noindent
whereas for $\alpha =\beta =\gamma =1$ it gives

$$[\{B_i^\xi ,B_j^\eta  \},B_k^\varepsilon ]=
2\varepsilon \delta_{jk}
\delta_{\varepsilon,-\eta}B_i^\xi +
2\varepsilon
\delta_{ik}\delta_{\varepsilon,-\xi}
B_j^\eta .  \eqno(10)  $$

Equations (9) and (10) are the defining relations for the
para-Fermi and para-Bose operators, respectively [3].

\bigskip
One can also say that the relations (8) define a structure of a
Lie-super triple system  on $G(n/m)$ [4]  with a triple product
$$
G(n/m)\otimes G(n/m)\otimes G(n/m) \rightarrow G(n/m)
$$
defined from (8).

\bigskip
It is important to point out that the relations (8)
define completely the supercommutation relations between all
generators. In order to show this
one has to use simply the (graded) Jacoby indentity:

$$
\[\[a,b\],\[c,d\]\]= \[\[\[a,b\],c\],d\]+
(-1)^{(deg(a)+deg(b))deg(c))}\[c,\[\[a,b\],d\]\].
$$

\noindent
Hence the triple relations (8) define completely the Lie
superalgebra $osp(2n+1/2m)$.

\bigskip
We have given a definition of $osp(2n+1/2m)$ in terms of a particular
$(2n+2m+1)\times (2n+2m+1)$ matrix realization. The triple
relations (8) are however representation independent.
\bigskip
In order to give a representation independent definition, denote
by $U$ the associative superalgebra with unity, (abstract) generators

$$ C_j^\pm (0) \equiv F_j^\pm, \quad j=1,2,\ldots,n \eqno(11)$$

\noindent
as even elements and

$$ C_i^\pm(1) \equiv B_i^\pm, \quad i=1,2,\ldots,m \eqno(12)$$

\noindent
as odd elements, which obey the triple relations (8).
The supercommutator $\[a,b\]$ between any homogeneous elements
$a, b$ is defined through the associative multiplication in $U$:

$$\[a,b\]=ab-(-1)^{deg(a)deg(b)}ba. \eqno(13)  $$

\noindent
With respect to the binary operation $\[\;\; ,\; \; \ \]$
$U$ is  a Lie superalgebra.

\smallskip
{\it Proposition 2.} The (finite-dimensional) subspace
$$
\eqalign{
& B(n/m)=lin.env.\{\[C_i^\xi (\alpha),C_j^\eta (\beta)\],\;
C_k^\varepsilon (\gamma)|
 \forall i,j,k; \;
\xi ,\eta ,\varepsilon =\pm ; \; \alpha ,\beta
,\gamma\in \Z_2 \} \cr
}\eqno(14)
$$

\noindent
of $U$ is a subalgebra of the Lie superalgebra $U$, isomorphic to
$osp(2n+1/2m)$. $U$ is the universal enveloping algebra
$U[osp(2n+1/2m)]$ of $osp(2n+1/2m)$.

\smallskip
Observe that the para-Bose (pB) operators neither commute nor
anticommute with the para-Fermi (pF) operators.

\smallskip
One can express any other generator (or, more generally, any other
element from $U$) in terms of the preoscillator generators (=
pB \& pF operators).

In particular, one can express the Chevalley generators of
$osp(2n+1/2m)$. To this end set:
$$
\eqalign{
& C_i^\pm(0)\equiv C_i^\pm,\quad i=1,2,\ldots, n, \cr
& C_j^\pm(1)\equiv C_{j+n}^\pm,\quad j=1,2,\ldots, m. \cr
}\eqno(15)
$$
\noindent
Then one has
$$
\eqalign{
& e_i={1\over 2}\[C_i^-,C_{i+1}^+\],\; i=1,\ldots,n+m-1,\cr
& e_{n+m}=-{1\over{\sqrt{2}}}C_{n+m}^- ,\cr
& f_i={1\over 2}\[C_{i+1}^-,C_{i}^+\],\; i=1,\ldots,n+m-1,\cr
& f_{n+m}={1\over{\sqrt{2}}}C_{n+m}^+ ,\cr
& h_i={1\over 2}\[C_{i+1}^+,C_{i+1}^-\]-\[C_{i}^+,C_{i}^-\],
  \; i=1,\ldots,n+m-1,\cr
& h_{n+m}=-{1\over 2}\[C_{n+m}^+,C_{n+m}^-\].  \cr
}\eqno(16)
$$

As is well known the Chevalley generators describe completely
the corresponding algebra, in this case $osp(2n+1/2m)$.
The preoscillator generators give an alternative description.
In terms of the latter it is easy to express various subalgebras
of $osp(2n+1/2m)$:
$$
\eqalign{
& osp(2n+1/2m)=lin.env.\{C_i^\xi,\; \[C_j^\eta,C_k^\epsilon \]
\vert i,j,k= 1 ,\ldots,n+m,\; \xi, \eta, \epsilon =\pm \};  \cr
} \eqno(17)
$$
\smallskip

$$
\eqalign{
& osp(2n/2m)=lin.env.\{ \[C_j^\eta,C_k^\epsilon \]
\vert j,k=1 ,\ldots,n+m,\;  \eta, \epsilon =\pm \};  \cr
}\eqno(18)
$$
\smallskip
$$
\eqalign{
& gl(n/m)=lin.env.\{ \[C_j^+,C_k^- \]
\vert j,k=1 ,\ldots,n+m \};  \cr
} \eqno(19)
$$
\smallskip
$$
\eqalign{
so(2n+1)&=lin.env.\{C_i^\xi,\; [C_j^\eta,C_k^\epsilon ]
\vert i,j,k=1 ,\ldots,n,\; \xi, \eta, \epsilon =\pm \}  \cr
&=lin.env.\{ F_i^\xi,\; [F_j^\eta,F_k^\epsilon ]
\vert i,j,k= 1 ,\ldots,n,\; \xi, \eta, \epsilon =\pm \};  \cr
}\eqno(20)
$$
\smallskip
$$
\eqalign{
& sp(2m)=lin.env.\{ \{C_j^\eta,C_k^\epsilon \}
\vert j,k=n+1 ,\ldots,n+m,\;  \eta, \epsilon =\pm \}  \cr
&\hskip 11mm =lin.env.\{ \{B_j^\eta,B_k^\epsilon \}
\vert j,k= 1 ,\ldots,m,\;  \eta, \epsilon =\pm \};  \cr
}\eqno(21)
$$
\smallskip
$$
\eqalign{
& gl(n)=lin.env.\{ [C_j^+,C_k^- ] \vert j,k=1 ,\ldots,n \}  \cr
& \hskip 9mm =lin.env.\{ [F_j^+,F_k^- ] \vert j,k=1 ,\ldots,n \};  \cr
} \eqno(22)
$$
\smallskip
$$
\eqalign{
& gl(m)=lin.env.\{ \{C_j^+,C_k^- \} \vert j,k=n+1 ,\ldots,n+m \}  \cr
& \hskip 9mm =lin.env.\{ \{B_j^+,B_k^- \} \vert j,k=1 ,\ldots,m \};  \cr
}\eqno(23)
$$
\smallskip
{\it Proposition 3} [1]. Let
$$
f_i^\pm, \; i=1,\ldots,n \quad {\rm be\; Fermi\; operators \; and}
$$
$$
b_j^\pm, \; j=1,\ldots,m \quad {\rm be\; Bose\; operators}
$$
\noindent
under the additional requirement that the Bose operators
{\bf anticommute} with the Fermi operators,
$$
\{f_i^\xi , b_j^\eta \}=0 \quad \forall i,j \;{\rm and}\;\xi, \eta.
$$
\noindent
Then the map
$$
F_i^\xi \; \rightarrow \; f_i^\xi ,\quad
 B_j^\eta \; \rightarrow \; b_j^\eta
$$
defines a representation of $osp(2n+1/2m)$.

\bigskip
This construction is rather unconvenional in the following sense:

\bigskip
1. The Bose operators are odd, fermionic variables and the Fermi
operators are even, bosonic variables.

2. The Bose and the Fermi operators anticommute.

\smallskip
\noindent
To the same conclusion arrived recently also Okubo [4] and Macfarlane [5].
\bigskip
In view of the above construction it is natural to ask questions like:

\smallskip
\noindent
$\bullet \;\; $ Can one define deformed preoscillator operators
and describe the quantum algebra $U_q[osp(2n+1/2m)]$ in terms of
these operators?
\smallskip
\noindent
$\bullet \;\; $ If yes, then what is the relation of these
operators to the deformed Bose and Fermi operators, introduced
in [6-9]?
\smallskip
\noindent
$\bullet \;\; $ How do the deformed preoscillator and the
deformed oscillator realizations look like?

\bigskip
The general answer to all of the above questions in unknown.
At present some of these questions can be answered  for
the deformed $so(2n+1)$ and $osp(1/2m)$ (super)algebras.

\bigskip
We proceed now to outline how this can be done first for the
odd-orthogonal Lie algebra $so(2n+1)$ for any $n$.

\bigskip\bigskip
\noindent
{\bf 2. Quantization of so(2n+1) with deformed para-Fermi operators [10]}
\bigskip
So far the algebra $so(2n+1)$ and more precisely - its universal
enveloping algebra $U[so(2n+1)]$ - has been quantized in terms
of its Chevalley generators. Let ${\hat e}_i, {\hat f}_i, h_i,\;
i=1,\ldots,n$ be the nondeformed Chevalley generators. Then
$U[so(2n+1)]$ is the associative unital (=with unity) algebra with
(free) generators
$${\hat e}_i,\; {\hat f}_i,\; h_i,\quad i=1,\ldots,n,$$

\noindent
which obey the Cartan relations
$$[h_i,h_j]=0,\quad
[h_i,{\hat e}_i]=\alpha_{ij}{\hat e}_i,
\quad [h_i,{\hat f}_i]=-\alpha_{ij}{\hat f}_i,\quad
[{\hat e}_i,{\hat f}]=\delta_{ij}h_i \eqno(24)
$$
\noindent
and the Serre relations
$$
[{\hat e}_i,{\hat e}_j]=0, \quad [{\hat f}_i,{\hat f}_j]=0,
\quad \vert i-j \vert>1,
$$
$$[{\hat e}_i,[{\hat e}_i,{\hat e}_{i \pm 1}]]=0
\;\;
[{\hat f}_i,[{\hat f}_i,{\hat f}_{i \pm 1}]]=0, \;\; i\neq n,\eqno(25)
$$
$$[{\hat e}_n,[{\hat e}_n,[{\hat e}_n,{\hat e}_{n-1}]]]
=0 , \;\;
[{\hat f}_n,[{\hat f}_n,[{\hat f}_n,{\hat f}_{n-1}]]]
=0.
$$

\smallskip
\noindent
The Cartan matrix $(\alpha_{ij})$ is
taken to be symmetric  with
$$\alpha_{nn}=1,\; \alpha_{ii}=2, \; i=1,\ldots,n-1,$$
$$\alpha_{j,j+1}=\alpha_{j+1,j}=-1, \; j=1,\ldots,n-1,\eqno(26) $$
\noindent
and all other $\alpha_{ij}=0$.
\bigskip
On the other hand, as we have alreay indicated,
$U[so(2n+1)]$ is the algebra of $n$-pairs of pF operators,
($\xi, \eta, \epsilon = \pm$ or $\pm 1$, $i,j,k=1,2,\ldots
,n)$:
$$
[[{\hat F}_i^\xi,{\hat F}_j^\eta ],{\hat F}_k^\epsilon]
= {1\over 2} (\epsilon - \eta)^2 \delta_{jk}{\hat F}_i^\xi
-{1\over 2} (\epsilon - \xi)^2 \delta_{ik}{\hat F}_j^\eta .
\eqno(27)
$$

The expressions of the Chevalley generators in terms of
the pF operators read:

$$
\vcenter{\openup3\jot \halign{$#$ \hfil & \hskip 4pt $#$ \hfil
 \cr
{\hat e}_n={1\over \sqrt 2}{\hat F}_n^-,
&{\hat e}_i={1\over 2}[{\hat F}_i^-,{\hat F}_{i+1}^+], \cr
{\hat f}_n={1\over \sqrt 2}{\hat F}_n^+,
&{\hat f}_i={1\over 2}[{\hat F}_{i+1}^-,{\hat F}_i^+], \cr
h_n={1\over 2}[{\hat F}_n^-,{\hat F}_n^+],
&h_i={1\over 2}[{\hat F}_i^-,{\hat F}_i^+]-
{1\over 2}[{\hat F}_{i+1}^-,{\hat F}_{i+1}^+] . \cr
}} \eqno(28)
$$

\smallskip
\noindent
The inverse relations are not that simple
($i=1,\ldots,n-1$):

$$\vcenter{\openup3\jot\halign {$#$ \hfil \cr
{\hat F}_i^-=\sqrt{2}
[{\hat e}_i,[{\hat e}_{i+1},[{\hat e}_{i+2},[\ldots,[{\hat
e}_{n-2},[{\hat e}_{n-1},{\hat e}_n]] \ldots ], \cr
{\hat F}_i^+=\sqrt{2}
[\ldots [{\hat f}_n,{\hat f}_{n-1}],{\hat f}_{n-2}],\ldots],{\hat
f}_{i+2}],{\hat f}_{i+1}],{\hat f}_i], \cr
{\hat F}_n^+=\sqrt{2}{\hat f}_n, \quad
{\hat F}_n^-=\sqrt{2}{\hat e}_n . \cr
}}\eqno(29)
$$
\bigskip
\noindent
Following Khoroshkin and Tolstoy [11] we define the deformed UEA
$$U_q[so(2n+1)]\equiv U_q$$ in terms of its Chevalley generators
as follows: $U_q$ is the (free unital) associative algebra with
Chevalley generators ($i=1,\ldots,n $ )

$$ e_i,\; f_i,\; k_i=q^{h_i},\; {\bar k}_i
\equiv k_i^{-1}=q^{-h_i},\eqno(30)
$$
\noindent
which satisfy the Cartan relations

$$ k_ik_i^{-1}=k_i^{-1}k_i=1, \; \; k_ik_j=k_jk_i,$$
$$ k_ie_j=q^{\alpha_{ij}}e_jk_i,\;\;k_if_j=q^{-\alpha_{ij}}f_jk_i,
\eqno(31)$$
$$ [e_i,f_j]=\delta_{ij}{{k_i-{\bar k}_i}\over{q-{\bar q}}}$$

\noindent
and the Serre relations (${\bar q}\equiv q^{-1}$)

$$
[e_i,e_j]=0, \quad [f_i,f_j]=0, \quad \vert i-j \vert>1,
$$
$$
[e_i,[e_i,e_{i \pm 1}]_{\bar q}]_q=0, \quad
[f_i,[f_i,f_{i \pm 1}]_{\bar q}]_q=0,\quad i\neq n, \eqno(32)
$$
$$
[e_n,[e_n,[e_n,e_{n-1}]_{\bar q}]]_q=0,
$$
$$
[f_n,[f_n,[f_n,f_{n-1}]_{\bar q}]]_q =0.
$$

\noindent
Here and throughout
$$[a,b]_{q^n}=ab-q^nba \eqno(33)$$
and it is assumed that the deformation parameter $q$ is any complex
number except
$$q=0,\; \; q=1, \; \; q^2=1.$$

\smallskip
\noindent
The deformed pF operators $F_i^{\pm}$ are defined as follows:

$$\vcenter{\openup3\jot\halign {$#$ \hfil \cr
F_i^-=\sqrt{2}
[e_i,[e_{i+1},[e_{i+2},[\ldots,[
e_{n-1},e_n]_{{\bar q}}
]_{{\bar q}}\ldots ]_{{\bar q}}, \cr
F_n^-=\sqrt{2}e_n , \cr
F_i^+=\sqrt{2}
[\ldots [f_n,f_{n-1}]_q,f_{n-2}]_q,\ldots]_q,
f_{i+1}]_q,f_i]_q, \cr
F_n^+=\sqrt{2}f_n. \cr
}}\eqno(34) $$

\bigskip

Let
$$L_i=k_ik_{i+1}\ldots k_n,\;\;i=1,\ldots,n.  $$
We call the operators
$$F_i^\pm,\;\; L_i \;\; i=1,\ldots,n  \eqno(35)$$
preoscillator generators of $U_q[so(2n+1)]$.

\bigskip
{\it Proposition 4.} The defining relations (31), (32) of
$U_q[so(2n+1)]$ in terms of its Chevalley generators (30)
hold if and only if the preoscillator generators (35) satisfy the
relations:

$$\vcenter{\openup3\jot \halign{$#$ \hfil & \hskip 48pt
\hfil $#$ \cr
L_iL_i^{-1}=L_i^{-1}L_i=1, \; \; L_iL_j=L_jL_i,  \cr
L_iF_j^{\pm}=q^{\mp\delta_{ij}}F_j^{\pm}L_i,  \quad i,j=1,\ldots,n,  \cr
[F_i^-,F_i^+]=2{L_i-{\bar L}_i\over q-{\bar q}},
\quad i=1,\ldots,n, \cr
[[F_i^{-\eta},F_{i\pm 1}^{\eta}],F_j^{-\eta}]_{q^{\pm
\delta _{ij}}}=2\delta _{j,i \pm 1}L_j^{\pm \eta}F_i^{-\eta},
\quad \eta = \pm, \cr
[F_n^\xi,[F_n^\xi,F_{n-1}^\xi]]_{\bar q}=0, \quad \xi=\pm.
 \cr
}} \eqno(36) $$
\bigskip
\noindent
Therefore $U_q[so(2n+1)]$ can be viewed as a free associative
unital algebra of the preoscillator generators with relations
(36).

\bigskip
In terms of the preoscillator generators it is very easy to
write an analogue of the Cartan-Weyl basis:

\bigskip

\noindent
{\it Proposition 5}.  The operators ($\xi=\pm$)
$$L_i,\; F_i^\pm,\;
[F_i^-,F_j^+],\;  [F_p^\xi,F_q^\xi],\; i \neq j,\;
i,j,p,q=1,\ldots,n $$
\noindent
are an analogue of the Cartan-Weyl
generators for $so(2n+1)$. In terms of these generators one
can introduce a basis in  $U_q[so(2n+1)]$.

\bigskip\bigskip
\noindent
{\bf 3. Quantization of osp(1/2n) with deformed para-Bose operators
}
\bigskip
\noindent
{\bf A. First realization [12-15]}

\bigskip

We proceed first to introduce $U_q\equiv U_q[osp(1/2n)]$ in
terms of its Chevalley generators.  The Cartan matrix
$(\alpha_{ij})$ is chosen as before, i.e., as a $n \times n $
symmetric matrix with
$$\alpha_{nn}=1,\; \alpha_{ii}=2, \;
i=1,\ldots,n-1,$$ $$\alpha_{j,j+1}=\alpha_{j+1,j}=-1, \;
j=1,\ldots,n-1,
$$
\noindent
and all other $\alpha_{ij}=0$.

Then $U_q$ is the associative superalgebra with Chevalley
generators
$$e_i,\; f_i,\; k_i=q^{h_i},\; i=1,\ldots,n,
$$
which satisfy the Cartan-Kac relations

$$
k_ik_i^{-1}=k_i^{-1}k_i=1, \quad k_ik_j=k_jk_i, \quad
 i,j=1,\ldots,n,
$$
$$
k_ie_j=q^{\alpha_{ij}}e_jk_i, \quad
k_if_j=q^{-\alpha_{ij}}f_jk_i, \quad  i,j=1,\ldots,n,
$$
$$
\{ e_n,f_n\} ={{k_n-k_n^{-1}}\over{q-q^{-1}}},\eqno(37)
$$
$$
[e_i,f_j]=\delta_{ij}{{k_i-k_i^{-1}}\over{q-q^{-1}}}
\quad \forall \; i,j  \; {\rm except} \;i=j=n,
$$
\smallskip
\noindent
the Serre relations for the simple positive root vectors
$$
[e_i,e_j]=0, \quad {\rm if} \quad i,j=1,\ldots,n \quad
{\rm and} \quad \vert i-j \vert >1,
$$
$$
e_i^2e_{i+1}-(q+q^{-1})e_ie_{i+1}e_i+e_{i+1}e_i^2=0,
   \quad i=1,\ldots,n-1,
$$
$$
e_i^2e_{i-1}-(q  +q^{-1})e_ie_{i-1}e_i+e_{i-1}e_i^2=0,
   \quad i=2,\ldots,n-1, \eqno(38)
$$
$$
e_n^3e_{n-1}+(1-q  -q^{-1})(e_n^2e_{n-1}e_n+e_ne_{n-1}e_n^2)+
e_{n-1}e_n^3=0,
$$

\noindent
and the Serre relations obtained from above by replacing
everywhere $e_i$ by $f_i$.

The grading on $U_q$ is induced from the requirement that the
generators $e_n,\; f_n$ are odd and all other generators are
even.
\bigskip
Passing to the para-Bose operators we first observe that the
nondeformed operators are defined with the relations
($\xi, \eta, \epsilon = \pm$ or $\pm 1$, $i,j,k=1,2,\ldots ,n$ )

$$
[\{{\hat A}_i^\xi,{\hat A}_j^\eta \},{\hat A}_k^\epsilon]
=(\epsilon - \xi)\delta_{ik}
{\hat A}_j^\eta + (\epsilon - \eta) \delta_{jk}{\hat A}_i^\xi.
\eqno(39)
$$
For the deformed pB operators we set:

$$
\vcenter{\openup3\jot\halign {$#$ \hfil \cr
A_i^-=-\sqrt{2}
[e_i,[e_{i+1},[\ldots,[e_{n-2},[e_{n-1},e_n]_{q^{-1}}
]_{q^{-1}}\ldots ]_{q^{-1}},
\quad i=1,\ldots, n-1), \cr
A_n^-=-\sqrt{2}e_n, \cr
A_i^+=\sqrt{2}
[\ldots [f_n,f_{n-1}]_{q},f_{n-2}]_{q},\ldots]_{q},
f_{i+1}]_{q},f_{i}]_{q},
\quad i=1,\ldots, n-1),  \cr
A_n^+=\sqrt{2}f_n. \cr
}}\eqno(40)
$$
\noindent
Introduce also $n$ even "Cartan" elements

$$
\eqalign{
& L_i=k_ik_{i+1} \ldots k_n=q^{H_i},\cr
& H_i=h_i + \ldots + h_n,
\quad i=1,\ldots,n.\cr
}\eqno(41)
$$
\bigskip
 The expressions of the Chevalley generators
in terms of the preoscillator generators read ($i\neq n$):

$$
\eqalign{
& e_n=-(2)^{-1/2}A_n^-,\quad
  e_i=-{q\over 2}\{A_i^-,A_{i+1}^+\}L_{i+1}^{-1},\cr
& f_n=(2)^{-1/2}A_n^+,  \quad
  f_i=-{1\over 2q}L_{i+1}\{A_i^+,A_{i+1}^-\} .\cr
}\eqno(42)
$$
\bigskip
\noindent
As in the pF case we call the operators
$$ L_i^{\pm 1}, \;A_i^\pm, \quad i=1,\ldots ,n \eqno(43) $$
preoscillator generators.

\bigskip
{\it Proposition 5.} The  relations of  $U_q[osp(1/2n)]$
in terms of its Chevalley generators
hold if and only if the preoscillator generators satisfy the
relations:

$$\vcenter{\openup3\jot \halign{$#$ \hfil & \hskip 48pt
\hfil $#$ \cr
L_iL_i^{-1}=L_i^{-1}L_i=1, \; \; L_iL_j=L_jL_i,  \cr
L_iA_j^{\pm}=q^{\mp\delta_{ij}}A_j^{\pm}L_i,  \quad i,j=1,\ldots,n,  \cr
\{A_i^-,A_i^+\}=-2{L_i-{\bar L}_i\over q-{\bar q}},
\quad i=1,\ldots,n, \cr
[\{A_i^{-\eta},A_{i\pm 1}^{\eta}\},A_j^{-\eta}]_{q^{\pm
\delta _{ij}}}=-2\eta \delta _{j,i \pm 1}L_j^{\pm \eta}A_i^{-\eta},
\quad \eta = \pm,  \cr
[\{A_{n-1}^\xi,A_n^\xi\},A_n^\xi]]_q=0, \quad \xi=\pm. &
\cr
}}\eqno(44)
$$
\bigskip
The relations (44) replace completely the Cartan-Kac relations (37)
and the Serre relations (38). They give an alternative definition of
$U_q[osp(1/2n)]$.
In terms of the preoscillator generators it is easy
to define all Cartan-Weyl generators:
$$ L_i^{\pm 1}, \;A_i^\pm,\;
\{A_i^-,A_j^+ \},\; \{A_i^\xi,A_j^\xi \},\quad i\neq j
=1,\ldots,n. \eqno(45) $$

\bigskip
Observe that the Cartan-Weyl root vectors are expressed in terms of
the pB operators exactly as in the nondeformed case.

In particular the operators
\smallskip
$$ L_i^{\pm 1}, \;
\{A_i^-,A_j^+ \}, \; i\neq j
=1,\ldots,n \eqno(46)$$

\smallskip
\noindent
give a realisation of the quantum $gl(n)$ algebra in terms of deformed
pB operators, which is exactly the same as in the nondeformed case.

\bigskip
Define a new set of operators
$$a_i^\pm, \; l_i={\hat q}^{N_i}\;i=1,\ldots,n \eqno(47)$$
where $q={\hat q}^2$, which satisfy the relations

$$
\eqalign{
& a_i^-a_i^+ -{\hat q}^{\pm 2}a_i^+a_i^-
  ={2\over {{\hat q}+{\hat q}^{-1}}}l_i^{\mp2},\;i=1,\ldots,n,\cr
&\cr
& a_i^\xi a_j^\eta ={\hat q}^{2\xi \eta}
  a_j^\eta a_i^\xi , \quad i<j,\quad
  \xi, \; \eta =\pm, \; i=1,\ldots,n.  \cr
}\eqno(48)
$$

\bigskip
For fixed $i$ the operators

$$
a_i^\pm, \; l_i={\hat q}^{N_i} \eqno(49)
$$
\smallskip
\noindent
are the same (up to multiple) as the deformed Bose CAOs [6-9].
The different modes however do not commute, they $q$-commute.

\bigskip
It is easy to check that the operators $a_i^\pm, \; l_i$ satisfy
the defining relations  (44) of the deformed algebra
$U_q[osp(1/2n)]$. Therefore we have
\bigskip
{\it Proposition 6.} The map
\smallskip
$$
A_i^\pm \rightarrow a_i^\pm \quad L_i \rightarrow
{\hat q}^{-1}l_i^{-2} \eqno(50)
$$
\smallskip
defines a representation, a kind of Fock representation of
$U_q[osp(1/2n)]$.

\bigskip

In particular the operators

\smallskip
$$ l_i, \;
\{a_i^-,a_j^+ \}, \; i\neq j
=1,\ldots,n \eqno(51)$$

\smallskip
\noindent
give a realisation, a Schwinger realisation of the quantum
$gl(n)$ algebra in terms of new kind of deformed Bose operators.
Elsewhere we will study the Fock representations of the creation and the
annihilation operators (48) and the related oscillator representations
of $U_q[osp(1/2n)]$ and $U_q[gl(n)]$.

\bigskip\bigskip
\noindent
{\bf B. Second realization. Quantization in terms of
Biedenharn-Macfarlane generators}
\bigskip
It may be more convenient to quantize
$U_q[osp(1/2n)]$ in terms
of new generators, namely $(i=1,\ldots,n)$
\bigskip
$$
\eqalign{
& B_i^-=\q^{n-i}\sqrt{{\q+\q^{-1}}\over{2\q}}A_i^-L_i^{-{1\over{2}}}
  L_{i+1}^{-1}L_{i+2}^{-1} \ldots L_n^{-1}, \cr
& B_i^+=\q^{i-n}\sqrt{{\q+\q^{-1}}\over{2\q}}A_i^+L_i^{1\over{2}}
  L_{i+1}L_{i+2}\ldots L_n, \cr
& K_i=\q^{-{1\over{2}}}L_i^{-{1\over{2}}}.\cr
}\eqno(52)
$$
\bigskip
{\it Proposition 7.}  $U_q[osp(1/2n)]$ is the (free unital) associative
algebra with generators $B_i^\pm$, $K_i$, $i=1,\ldots,n$ and the relations
$$\vcenter{\openup3\jot \halign{$#$ \hfil & \hskip 48pt
\hfil $#$ \cr
K_iK_i^{-1}=K_i^{-1}K_i=1, \; \; K_iK_j=K_jK_i,  \cr
K_iB_j^{\pm}=\q^{\pm\delta_{ij}}B_j^{\pm}K_i,  \quad i,j=1,\ldots,n,  \cr
\{B_i^-,B_i^+\}={{\q K_i^2-q^{-1}K_i^{-2}}\over{{\hat q} -{\hat q}^{-1}}},
\quad i=1,\ldots,n, \cr
[\{B_{i\pm 1}^{-\eta},B_{i}^{\eta}\}_{{\q^{\mp 2}}},
B_j^{-\eta}]_{\q^{\pm 2\delta_{ij}}}=-\eta(1+\q^{\pm 2\eta})
\delta _{ij}K_j^{\pm 2\eta}B_{i\pm 1}^{-\eta},
\quad \eta = \pm,  \cr
[\{B_{n-1}^\xi,B_n^\xi\}_{{\q}^2},B_n^\xi]]=0, \quad \xi=\pm. &\cr
}}\eqno(53)
$$
\bigskip
The reason to introduce the operators (52) stems from the next
proposition.

\bigskip
{\it Proposition 8.} Let $b_i^\pm, \; k_i$ be the
deformed CAOs with $k_i=\q^{N_i}$ [6-9]:
\smallskip

$$
\eqalign{
& b_i^- b_i^+ -{\q}^{\pm2}b_i^+b_i^-=k_i^{\mp2},\cr
& k_ib_i^\pm=\q^{\pm}b_i^{\pm}k_i \cr
}\eqno(54)
$$
\smallskip
\noindent
Assume that the different modes of such
operators commute. Then the map

$$
B_i^\pm \rightarrow b_i^\pm, \quad K_i \rightarrow k_i \eqno(55)
$$
\smallskip
\noindent
defines a representation of  $U_q[osp(1/2n)]$.
\smallskip
We call the generators (52) Biedenharn-Macfarlane generators of
$U_q[osp(1/2n)]$.  According to Proposition 7 the
Biedenharn-Macfarlane generators give an alternative definition
of the deformed orthosymplectic superalgebra $U_q[osp(1/2n)]$.
The main algebraic feature of these generators stems from
Proposition 8: in the Fock representation they coincide with the
deformed creation and annihilation operator (54).

\smallskip
It is not easy to write down all triple relations the
Biedenharn-Macfarlane  operators satisfy.
Here are some of them:
$$\vcenter{\openup3\jot \halign{$#$ \hfil & \hskip 48pt
\hfil $#$ \cr
[\{B_i^-,B_j^+\}_{\q^2},B_k^+]=0, \quad i<j\leq k \;{\rm or} \;k<i<j, \cr
[\{B_i^-,B_j^+\}_{\q^2},B_k^+]_{\q^4}=(1-\q^4)
\{B_i^-,B_k^+\}_{\q^2}B_j^+ , \quad  i<k<j,  \cr
[\{B_i^-,B_j^+\}_{\q^2},B_i^+]_{\q^2}=(1+\q^2)
B^+_j K_i^{-2}, \; i<j, \cr
[\{B_i^-,B_j^+\}_{\q^2},B_k^-]=0, \quad k\leq i<j \;{\rm or} \; i<j<k, \cr
[\{B_i^-,B_j^+\}_{\q^2},B_k^-]_{\q^{-4}}=(1-\q^{-4})
\{B_k^-,B_j^+\}_{\q^2}B_i^+, \quad  i<k<j,  \cr
[\{B_i^-,B_j^+\}_{\q^2},B_j^-]_{\q^{-2}}=-(1+\q^{-2})
B^-_i K_j^{-2}, \; i<j, \cr
}}\eqno(56)
$$
\noindent
and the conjugate to the above relations:

$$\vcenter{\openup3\jot \halign{$#$ \hfil & \hskip 48pt
\hfil $#$ \cr
[\{B_i^-,B_j^+\}_{\q^{-2}},B_k^-]=0, \quad j<i\leq k \quad{\rm or}
\quad k<j<i ,\cr
[\{B_i^-,B_j^+\}_{\q^{-2}},B_k^-]_{\q^4}=(1-\q^4)
\{B_k^-,B_j^+\}_{\q^{-2}}B_i^- ,\quad j<k<i , \cr
[\{B_i^-,B_j^+\}_{\q^{-2}},B_j^-]_{\q^2}=-(1+\q^2)
B^-_i K_j^{2}, \quad i>j, \cr
[\{B_i^-,B_j^+\}_{\q^{-2}},B_k^+]=0,\quad k\leq j<i
\quad {\rm or} \quad j<i<k, \cr
[\{B_i^-,B_j^+\}_{\q^{-2}},B_k^+]_{\q^{-4}}=(1-\q^{-4})
\{B_i^-,B_k^+\}_{\q^{-2}}B_j^+ ,\quad j<k<i,  \cr
[\{B_i^-,B_j^+\}_{\q^{-2}},B_i^+]_{\q^{-2}}=(1+\q^{-2})
B^+_j K_i^{2}, \quad j<i. \cr
}}\eqno(57)
$$

\bigskip
Using the above relations we will show elsewhere that all
operators $k_i$ together with

$$
\eqalign{
& \{b_i^-,b_j^+\}_{\q^2}\quad {\rm for}\quad i<j ,\cr
& \{b_i^-,b_j^+\}_{\q^{-2}}\quad {\rm for}\quad i>j
}\eqno(58)
$$
\smallskip
\noindent
define a representation of Cartan-Weyl generators of
$U_q[gl(n)]$ and we will study the corresponding oscillator
representations.
\bigskip\bigskip

4. {\bf Comments}
\bigskip
We have not touched here the coalgebra and, more generally, the entire
Hopf algebra structure of $U_q[so(2n+1)]$ and $U_q[osp(1/2n)]$. One can
in principle derive the expression for the action of the comultiplication
$\Delta$ on the creation and the annihilation operators using the
known relations for the action of $\Delta$ on the Chevalley generator and
the expressions for the preoscillator generators  in terms of the Chevalley
operators. The corresponding expressions are however extremely involved
and we do not have any  explicit formulae for $\Delta(B_i^\pm)$ for instance.
On the other hand one needs such relations in order
to give a complete Hopf algebra description of
$U_q[so(2n+1)]$ and $U_q[osp(1/2n)]$ in terms of preoscillator
generators. Therefore we will conclude the present talk with a problem.
\smallskip
{\bf Problem.} Find closed expressions for the action of the
comultiplication and the antipode on the preoscillator generators.

\bigskip\bigskip
5. {\bf Acknowledgments}
\bigskip

It is a pleasure to thank Dr. J. Van der Jeugt, Dr. R.
Jaganathan and Dr. N. I. Stoilova for the stimulating
discussions. The author is thankful to Prof. Guido Vanden Berghe
for the kind hospitality at the Department of Applied
Mathematics and Computer Science, University of Ghent.

This work was supported by a grant awarded by the Belgian
National Fund for Scientific Research and by the contract
$\Phi$-416 of the Committee of Science of Bulgaria.

\vskip 12mm
{\bf References}
\bigskip

{\settabs\+[11] & I. Patera, T. D. Palev, Theoretical interpretation of the
   experiments on the elastic \cr

\+ [1] & T. D. Palev, Journ. Math. Phys. {\bf 23} (1982) 1100. \cr
\+ [2] & V. G. Kac, Lect. Notes in Math. {\bf 626} (1978) 597. \cr
\+ [3] & H. S. Green, Phys. Rev. {\bf 90} (1953) 270. \cr
\+ [4] & S. Okubo Journ. Math. Phys. {\bf 35} (1994) 2785 \cr
\+ [5] & A. J. Macfarlane, Preprint DAMPT 93-37 (1993).\cr
\+ [6] & A. J. Macfarlane, Journ. Phys. A {\bf 22} (1989) 4581.\cr
\+ [7] & L. C. Biedenharn, Journ. Phys. A {\bf 22} (1989) L873.\cr
\+ [8] & C. P. Sun and H. C. Fu, Journ. Phys. A {\bf 22} (1989) L983.\cr
\+ [9] & T. Hayashi, Comm. Math. Phys. {\bf 127} (1990) 129.\cr
\+ [10] & T. D. Palev, Lett. Math. Phys. {\bf 31} (1994) 151 \cr
\+ [11] & S. M. Khoroshkin and V. N. Tolstoy, Comm. Math. Phys.
          {\bf 141} (1991) 599.\cr
\+ [12] & E. Celeghini, T. D. Palev and M. Tarlini,
          Mod. Phys. Lett. B {\bf 5} (1991) 187.\cr
\+ [13] & T. D. Palev and N. I. Stoilova, Lett. Math. Phys.
          {\bf 28} (1993) 187. 	\cr
\+ [14] & T. D. Palev, Lett. Math. Phys.
          {\bf 28} (1993) 321. \cr
\+ [15] & T. D. Palev, J. Phys. A {\bf 26} (1993) L1111.\cr

\end